\documentclass[prl,twocolumn,floats,aps,superscriptaddress,showpacs]{revtex4}

\usepackage{amsmath}
\usepackage{epsfig}

\newcommand{\bv}{{\bf  v}}
\newcommand{\bu}{{\bf  u}}
\newcommand{\bx}{{\bf  x}}

\newcommand{\br}{{\bf  r}}

\newcommand{\BE}{\begin{equation}}
\newcommand{\EE}{\end{equation}}
\newcommand{\BA}{\begin{array}}
\newcommand{\EA}{\end{array}}
\newcommand{\beqn}{\begin{eqnarray}}
\newcommand{\eeqn}{\end{eqnarray}}

\newcommand{\etab}{\mbox{\boldmath $\eta$}}

\begin{document}

\title{Self-propelled  non-linearly diffusing particles.
Aggregation and continuum description.
}

\author{Crist\'obal L\'opez}
\affiliation{Instituto
Mediterr\'aneo de Estudios Avanzados IMEDEA (CSIC-UIB),
     Campus de la Universitad de las Islas  Baleares,
E-07122 Palma de Mallorca, Spain.
} 

\date{\today}

\begin{abstract}

We introduce a  model of self-propelled particles   carrying out  a Brownian motion
with a diffusion coefficient which depends on the local density of particles within
a certain finite radius.
Numerical simulations  show
that in a range of parameters the long-time spatial distribution of particles is non-homogeneous,
 and clusters  can be observed.
A number density equation,  which explains
the emergence of the aggregates and indicates the values of the parameters for which they appear, is derived.
Numerical results  of this continuum density equation are also shown.

\end{abstract}
\maketitle




The tendency of living individuals  to aggregate is an ubiquitous phenomenon in Nature that
constitutes a central problem in different areas of the natural sciences. 
The usually discussed examples in the literature
arise from very diverse contexts and comprise a  wide range of  spatiotemporal
scales: human crowds, animal groups, cell populations or bacteria colonies.
In this context distinct situations are studied, including   the dynamics of groups of
individuals moving coherently resembling
the behavior
of fish schools, bird flocks or insect swarms \cite{flierl},
 the pattern formation in populations of bacteria \cite{murray},
or the patchy structure of plankton mediated by hydrodynamic driving \cite{nuestroreview}.
Two levels of descriptions
are usually performed, one considering the discrete particle dynamics, and the other 
accounting for the large spatiotemporal scales 
in terms of an evolution equation for the number density of particles.
In the best of the cases, the last is
derived  from the more fundamental particle dynamics.

From the theoretical point of view one of the main questions one can address is about what causes
aggregation to form \cite{keshet}. 
In order to answer this, and starting from the discrete particle dynamics,
 many different mathematical
models have been introduced 
which, mainly in the physics literature, are characterised by their simplicity but also because  they retain
the basic features underlying the clustering phenomenon. 
Essentially these models assume that
i) individuals are self-propelled, 
and ii) they interact with the {\it environment} and/or with other particles through attractive and repulsive forces
\cite{flierl}, or rather, by modifying its velocity because of communications with neighbours \cite{VM}. 
Note, however, that even a simpler  mechanism that cannot be classified within ii) and that gives rise
to clustering  has been
introduced in \cite{young}.

In this letter, following this line of searching simplicity,
we introduce a basic model 
going through steps i) and ii) characterised because 
the aggregation phenomenon
is due to a new type of  mechanism, where no attractive nor repulsive forces among particles are taken into account.
The model considers  a fix population of particles
moving Brownianly. At every time step, any particle modifies, reducing, its random motility (diffusion coefficient)
depending on the total
number of particles in a neighbourhood of spatial range $R$.  This response of the particles to their local environment
can be interpreted as an effective deceleration because of the presence (via collisions, interchange  of chemicals,
excluded volume effects, demographic pressure, etc...)
of other particles.
The interaction radius $R$ appears frequently in the modelling of biological 
systems to take into account
visual or hearing stimuli, chemical signaling, and other kinds of interactions at a distance.
$R$ is considered to be a fixed number in this work but in some biological species
it depends on the environmental conditions.
Just 
as an example, we mention the detection distance of preys  for a type of zooplankton \cite{jpr}.
In our model,
despite the particles are moving randomly and no attractive forces among them are considered,
they tend to cluster. It will be obvious, when the coarse-grained number
density equation of the model is presented,   that this is because of the nonlinear
character of the diffusion of the particles.
Therefore,
the aim of this work is to introduce this
new type of mechanism, the local modification of the motility by the crowding of the surroundings,
 at the level of discrete particle dynamics,
and show under which conditions it gives rise to clustering. Then obtain its continuum  description and interpret the model
and its clustering instability in terms of this. The manuscript goes precisely along these lines:
first we introduce the particle model and study
it numerically, then we derive its continuum description and study it analytically and numerically.

Let us consider $N$ pointwise particles initially distributed randomly
in a two-dimensional system of size $L \times L$  with periodic boundary conditions.
At every discrete time step $t$ the positions of all the particles, ${\bf x}_i(t)=(x_i(t),y_i(t))$ ($i=1,...,N$),
are updated synchronously as follows:
\begin{eqnarray}
x_i(t+\Delta t)= x_i(t) + \sqrt{\frac{2 D_0 \Delta t}{(N_R(i))^p}} \eta^x_i (t), \nonumber \\
y_i(t+\Delta t)= y_i(t) + \sqrt{\frac{2 D_0 \Delta t}{(N_R(i))^p}} \eta^y_i (t), 
\label{modelo}
\end{eqnarray}
where $D_0$ is a constant named in the following constant  diffusivity or motility (this last name
is the usual one  for cells, bacteria or plankton organisms),
 $\Delta t$ is the
time step which we take equals to $1$ in the numerical simulations, $\etab=(\eta^x, \eta^y)$ is a white noise
with zero mean and correlations $<\eta^a_i (t), \eta^b_j (t')>= \delta_{ab} \delta_{ij} \delta_{t t'}$.
$N_R(i)$ denotes  the total number of particles at distance less than $R$ of particle $i$ including itself, and
$p$ is a positive real number.
The model clearly states that particles are moving Brownianly with diffusion coefficient inversely proportional to
the number of neighbours within a range $R$,
$D_i=D_0/(N_R(i))^p$, being $D_i$  the {\it effective}
diffusivity of particle $i$. In this way,
if the neighbourhood of a particle is poorly crowded it diffuses quickly.
The functional dependence of $D_i$  resembles
the standard models of density-dependent
dispersal of insects or animals \cite{murray}, where
the {\it behavioral} parameter $p$ determines this dependence. 
Note, however, that in 
in these models, at difference with the one introduced in this work,
the diffusivity increases with
the density of particles. 
A distinct type of decreasing function of $D_i$ with $N_R(i)$ could be considered,
and will be discussed later on at the
level of the density equation.

First, we pay attention to the limits $R \to 0$ and $R \to L$ (because of periodic boundary conditions the total system
size limit for $R$ is in fact smaller than $L$ but we keep this notation just for clarity).
In the first one, $N_R(i)=1 $ for all $i$ so that
the particles  diffuse with the constant motility, $D_i=D_0$. In the second limit again all the particles diffuse with
the same diffusion coefficient given by $D_i =D_0/N$, that is negligible for large $N$.
Of course we do not expect any grouping behavior in
these two limits and $R$ will be considered in the rest of the paper to be in the interval  $0< R < L$. 
In addition, for large values of 
 $p$   the diffusivity of the particles is almost zero, so we neither consider them. 
Similarly, very small values of $D_0$ are disregarded.

\begin{figure}
\epsfig{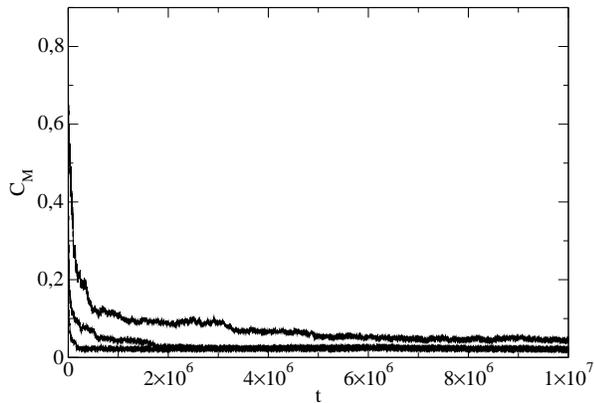}
\caption{We see in this figure the approach to the steady state. 
We plot the clustering coefficient, which gives an idea of the spatial distribution of particles (see text),
vs time for different values of $D_0$ and the same value of $p=2$. From top to bottom:
 $D_0=10^{-5}$,  $D_0=10^{-4}$,  $D_0=10^{-3}$. The smaller $D_0$ the larger the time needed to reach the steady state.
}
\label{fig:stationary}
\end{figure}

Numerical simulations of eq.~(\ref{modelo}) show that a statistical steady state of the particles distribution is reached.
We observe this by computing the cluster coefficient  of the distribution of particles (see below), and observing
that for long-times it goes to an average
 constant value, which indicates that the spatial distribution of particles is statistically stationary
(see fig.~(\ref{fig:stationary})).
For the statistical stationary state we also note that: 
 a) for a wide range of values of $D_0$ and $R$  the long-time spatial distribution of particles is homogeneous if 
$p$ is in the range $0< p \leq 1$; b) for  $p > 1$ the particles distribute non-uniformly for all the values of $D_0$ and $R$ considered. 
Increasing the value of $p$ upon the 
critical value $p_c=1$ the
unhomogeneous distribution of particles turns into clusters.
However, if we keep on increasing $p$ (as already suggested in the former paragraph)
the effective diffusivity, $D_i$,  is almost zero and the particles remain 
almost without moving, so that the time scales to observe a
 nonhomogenous distribution become very large.
The aggregation proceeds
as follows: if a number  particles are close to each other their  effective diffusivity diminishes and
they do not
get much appart forming a small cluster. As other particles, which are moving
randomly in the system, get near  this cluster, their diffusivity
is reduced and they stay in the aggregate. 

\begin{figure}
\begin{center}
\epsfig{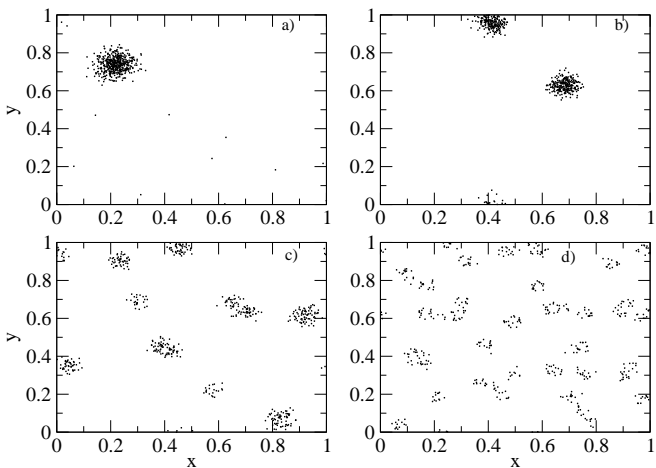}
\end{center}
\caption{Long-time ($10^7$ time steps)
 spatial distribution of particles. The parameters are $R=0.05$, initial density $\rho_0=500$ for all the plots,  and
a) $D_0=10^{-4}$, $p=1.5$;
b) $D_0=10^{-5}$, $p=2.0$;
c) $D_0=10^{-5}$, $p=3.5$;
d) $D_0=10^{-3}$, $p=7.5$. 
The system size is $L=1$.
}
\label{fig:patterns}
\end{figure}

No cluster dynamics is observed in the system.
The groups of particles  remain statistically
fixed in the space, not changing notably
their position, since most of the particles in any of them have
a very small diffusivity. Only the particles in the outer parts have a  non-negligible diffusion coeficient and slightly move, never
getting far of the cluster and only very rarely leaving the cluster and joining to another one.
In principle, it could be surprising that large values of $D_0$ do not avoid any possibility of aggregation in the system.
In fact, in the numerical simulations one observes
that larger values of $D_0$ favour encounters among particles 
and the clustered steady state is reached in a shorter time. 
Thus, the value of $D_0$ in the model only changes the time scale of the system.
As already mentioned,
increasing values of $p$ homogenises the spatial distribution of particles, i.e. they
are more sparsed in the system, and the number of groups increases.
When clusters are formed in the system, 
their typical size   strongly depends on the values of the parameters
 $R$,  $p$ and the initial number of particles. Also, it is always observed that the clusters 
do not periodically distribute in space \cite{nostropre} which is what is expected for  realistics models of swarming since periodicity
is rarely seen in these \cite{mogilnerkeshet}.
In fig.~(\ref{fig:patterns}) we show  distinct spatial structures observed for different
values of the parameters.
A single cluster and  several clusters
are shown there. 

\begin{figure}
\begin{center}
\epsfig{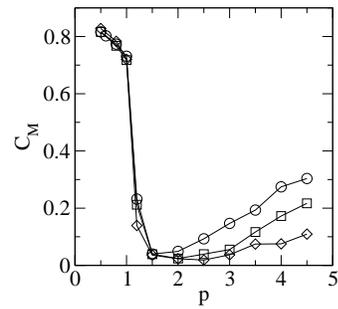}
\end{center}
\caption{Clustering coefficient versus the parameter $p$.
Each plot corresponds to a fixed value of $D_0$. The  circles
correspond to 
$D_0 =10^{-5}$, squares to  $D_0=10^{-4}$, and diamonds to $D_0=10^{-3}$.
The rest of the parameters are $R=0.05$ and $\rho_0=500$.
The system is divided in a number of $M=144$ boxes and every point in this figure
is calculated from a spatial distribution of particles at time $t=10^7$.
}
\label{fig:clustering}
\end{figure}

Clustering is quantified by using
 an  entropy-like measure 
$S_M=-\sum_{i=1}^M \frac{m_i}{N}\ln \frac{m_i}{N}$,
where $M$ is the number of boxes in which we divide the system, and $m_i$ is the number of particles
inside box $i$. One has that
$0\le S_M \le \ln M$, so that the minimum value, $S_M=0$, is obtained when all the particles are
in just one of the boxes, and the maximum one, $\ln M$,  is reached
when $m_i= N/M$ for all $i$, 
i.e., $S_M$ decreases when  there is more aggregation. The
clustering coefficient is defined as $C_M=\exp(<S_M>_t)/ M$, where $<.>_t$ denotes a temporal average in steady
state conditions,
so that when there is no clustering $C_M \approx 1$ and small values of $C_M$ indicate a non-uniform distribution of particles.
In fig.~(\ref{fig:clustering}) we plot $C_M$ versus $p$  for
different values of $D_0$, and always a    fixed $R=0.05$ (similar plots, not shown,
are obtained for other $R$'s).
The transition at $p_c=1$ is clearly observed for the different values of  $D_0$.
In addition, for a fixed $D_0$ one can also see
that as $p$ increases (for values in the clustering phase, $p>p_c$) $C_M$ also increases.
This indicates that the number of clusters in the system gets larger as $p$  increases,
as can be observed in fig.~\ref{fig:patterns}b) and c).

In order to obtain a continuum description of the model to try to understand better its properties,
 we first take the limit $\Delta t \to 0$ in eq.~(\ref{modelo})
and get the Langevin equation 
\begin{equation}
d{\bf x}_i (t)= \beta ({\bf x}_i (t),t) d{\bf W}_i(t), \ \  i=1,...,N,
\label{langevin}
\end{equation}
where $d{\bf W}_i(t)$ is a Wiener proccess and $\beta ({\bf x}_i (t),t)= \sqrt{2 D_0/N_R(i)^p}$. 
It is very important to note that eq.~(\ref{langevin}) should be
interpreted in the Ito Calculus. This is so because it represents 
the continuous time limit of a discrete population dynamics model of nonoverlapping 
generations. A clear discussion on this can be consulted in
the first of the references in \cite{vankampen}.

Then the mean-field number density equation for  eq.~(\ref{langevin}) (particularly appropriate for 
not too small densities)
 is given by \cite{vankampen}
\begin{equation}
\frac{\partial \rho(\bx,t)}{\partial t}=\frac{1}{2} \nabla^2 (\beta (\bx, t)^2 \rho(\bx,t)),
\label{fp}
\end{equation}
where $\rho(\bx,t)$ is the number density of particles in the continuum space-time.
Taking into account that in the continuum limit  $N_R(i)=N_R(\bx)=\int_{|\bx-\br|\leq R} d\br \rho(\br,t)$, the final
evolution equation for the density  in our model is
\begin{equation}
\frac{\partial \rho(\bx,t)}{\partial t}=D_0 \nabla^2 \left( \frac{\rho(\bx,t)}{\left(\int_{|\bx-\br|\leq R}
 d\br \rho(\br,t)\right)^p} \right),
\label{modelocontinuo}
\end{equation}
with initial condition $\rho(\bx,0) =N/L^2$. 
A proper derivation of eq.~(\ref{modelocontinuo}) from the interacting particle dynamics, 
eq.~(\ref{langevin}), gives rise to a multiplicative noise term that has been averaged out 
in eq.~(\ref{modelocontinuo}). As already mentioned, this is a good approximation for  not too
small densities. Moreover,  fluctuations in the density equation, which reflect the discrete nature
of the particles,
seem to  have  an irrelevant role in the pattern formation instability to be discussed below 
(see also \cite{nostropre}).

Particle number is conserved in eq.~(\ref{modelocontinuo}) and also it is implicitly assumed that when 
the density at a point is zero within a range $R$ it remains zero (any possible singularity is avoided in
eq.~(\ref{modelocontinuo})). In this density equation it is clear that the system is described by a 
nonlinear diffusion equation where the effective diffusivity at any point is decreased by the total
density in a neighbourhood of the point. With respect to other existent nonlinear diffusion models,  e.g. for
insect swarming \cite{murray},
our model presents  the following crucial differences: the  finite range of interaction $R$, 
and that, as has already been mentioned, in those {\it classical}
models the motility is generally  proportional to the local density, at least to the author best knowledge.
The following non-dimensionalization $s= D_0 t/R^2$, ${\bf u}={\bf x}/R$ and $\bar \rho =R^2 \rho$ transforms
eq.(\ref{modelocontinuo}) into 
\begin{equation}
\frac{\partial \bar \rho({\bf u},s)}{\partial s}= \nabla_{\bf u}^2 \left( \frac{\bar \rho({\bf u},s)}{\left(\int_{|\bu-\bv|\leq 1}
 d\bv \bar \rho(\bv,t)\right)^p} \right),
\label{modelocontinuonon}
\end{equation}
from which  it is clear that $D_0$  only {\it renormalizes} the time-scale of the system.

\begin{figure}
\epsfig{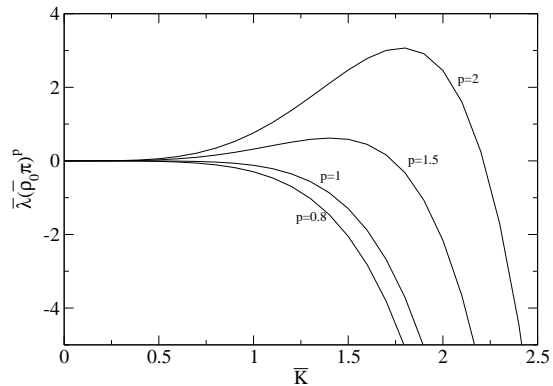}
\caption{Dispersion relation, eq.~(\ref{dispersion}), for different values of $p$.
In the vertical we plot the (non-dimensional) growth exponent $\bar \lambda$ multiplied by 
$(\bar \rho_0 \pi)^p$,
 and in the horizontal axis $\bar K = KR$.  $p=1$ is the
critical value for the instability.
}
\label{fig:dispersion}
\end{figure}

To study the clustering, let us  make a linear stability analysis of  the stationary homogeneous
solution, $\rho_0=N/L^2$, of eq.~(\ref{modelocontinuo}) . We  write
$\rho (\bx, t)=\rho_0 + \epsilon \psi (\bx, t)$ where $\epsilon$ is a small parameter,
and $\psi (\bx, t)$ the space-time dependent perturbation, and substitute it in
eq.~(\ref{modelocontinuo}). To first order in $\epsilon$, $\psi$ evolves as \cite{stratonovich}:
\begin{eqnarray}
\partial_t  \psi (\bx, t) = \frac{D_0}{(\rho_0 \pi R^2)^p} \nabla^2 \psi- \nonumber \\
\frac{D_0 p}{\rho_0^p (\pi R^2)^{p+1}} \nabla^2 \left( \int_{|\bx -\br | \leq R}\psi (\br, t)\, d\br \right).
\label{perturbacion}
\end{eqnarray} 
Then taking an harmonic perturbation $\psi (\bx, t)= \exp(\lambda t + i {\bf k} \cdot \bx)$
one arrives at the following dispersion relation
\begin{equation}
\lambda(K) = \frac{D_0 K^2}{(\rho_0 \pi R^2 )^p} (\frac{2 p J_1(KR)}{KR}-1),
\label{dispersion}
\end{equation}
being $K$ the modulus of ${\bf k}$, and $J_1$  the first-order
Bessel function.
(Non-dimensionalising eq.~(\ref{dispersion}), or rather performing the stability analysis directly to
the non-dimensional expression eq.~(\ref{modelocontinuonon}), one has that
$\bar \lambda(\bar K) =  (\bar K^2/(\bar \rho_0 \pi )^p) (\frac{2 p J_1(\bar K)}{\bar K}-1)$,
where $\bar K= KR$ and $\bar \lambda =\lambda R^2 /D_0$.)
 The uniform density is unstable, giving rise to aggregation, if
$\lambda $ is positive. For $K=0$ one has that  $\lambda =0$ for all $p$ (and any $D_0$, $\rho_0 $
and $R$). This implies
that the homogeneous density is neutrally stable to uniform perturbations which is due
to the conservation of the total number of particles. 
As shown in fig.~(\ref{fig:dispersion}) the growth rate $\lambda$ (always a real number)
is positive for $p >1$, while the system
is stable for $p\leq 1$. This is in agreement with the numerical results found for the discrete model.
Moreover, the instability for values of $p$ larger than $1$ is in a band of wave vectors
within the range $0 \leq K \leq K_+$, i.e., it is of type II in the classification of \cite{crosshohenberg}.
This kind of instability appears typically for systems with a conservation law, and are characterised by the fact that
the growth rate,  $\lambda$, vanishes at K=0, and 
represents the onset of aggregation and formation of groups~\cite{mogilnerkeshet}.
The phenomenology of aggregation already 
observed in the particle model has its perfect counterpart in  the dispersion relation
eq.~(\ref{dispersion}), being  the values of $D_0$, $\rho_0$ and $R$  irrelevant for the onset of
clustering.
Therefore, the density equation perfectly explains  the aggregation of the particles as a deterministic instability
of type II. The nonlinearities of the model saturate the exponential linear growth for $p > 1$ given by 
the dispersion relation.
It is worth mentioning that 
if an exponentially decreasing dependence fo the diffusivity would have been considered,
$D(i) \propto \exp{-(p N_R(i))}$, a similar relation dispersion is obtained
but with a control parameter $\tilde p$
dependent on $R$ and $\rho_0$, which is also rather realistic.

\begin{figure}
\epsfig{figure=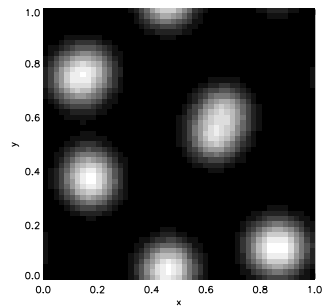,height=0.25\textwidth,clip=true,width=0.35\textwidth,angle=0}
\caption{Density distribution at large time. $D_0=10^{-3}$, $R=0.1$, $\rho_0=76.23$
and $p=2$. Lighter colour indicates larger density.
}
\label{fig:continua}
\end{figure}

We have also performed  a numerical simulation of the  density equation eq.~(\ref{modelocontinuo}) starting
from a random initial distribution and using a variable time step in order to avoid any numerical instability.
A long-time non-uniform non-periodic pattern
  is plotted in fig.~(\ref{fig:continua}) for $D_0=10^{-3}$, $R=0.1$, $\rho_0=76.23$ 
and $p=2$. As expected, but not shown, if $p \leq 1$ a uniform distribution of the field is
 obtained for these and other parameter values.

In summary, a new mechanism for clustering of self-propelled particles has been presented.
It just considers that the motility of any particle decreases with the 
crowding of its surroundings. 
The model has been studied and its 
number density equation  has been obtained, which turns out to be a particular form of 
a nonlinear diffusion equation.
The relation dispersion indicating the conditions for clustering were calculated, so that 
the clustering at the particle level can be understood as a deterministic instability of the
density equation. In addition, the role of the parameters of the model becomes clearer
at the density description level.
Regarding a biological application, 
motile bacteria or plankton particles  propel themself, so that they are 
typically modelled as
self-propelled particles, and maybe a minimal mechanism like the one
discussed in this work can explain some of the yet unknown causes for their clustering.
Ongoing research on the nature of the transition to clustering \cite{VM,chate},  on other possible 
functional shapes
of the motility of the particles, and on the influence of a variable interaction range
are in consideration for the next future.

I acknowledge a very useful conversation with Emilio Hern\'andez-Garc\'\i a
and also a critical reading of the manuscript.
Financial support from MEC (Spain) and FEDER through project
CONOCE2 (FIS2004-00953) is greatly acknowledged. C.L. is a {\sl
Ram{\'o}n y Cajal} fellow of the Spanish MEC.



\begin{thebibliography}{99}
\bibitem{stratonovich} If we would have taken, wrongly, Eq.~(\ref{modelocontinuo}) in its
Stratonovich sense, the evolution equation for the perturbation is given by just the diffusion term.
No instability is then observed  in this wrong framework.
\end{thebibliography}
\end{document}